\documentclass[%
 aip,
 jap,%
 amsmath,amssymb,
preprint,%
author-numerical,%
showkeys]{revtex4-1}

\usepackage{graphicx}
\usepackage{dcolumn}
\usepackage{bm}
\usepackage[colorlinks=true,allcolors=blue,breaklinks=true]{hyperref}

\begin{document}


\title{Critical shell thickness for InAs-Al\texorpdfstring{$_x$}{x}In\texorpdfstring{$_{1-x}$}{1-x}As(P) core-shell nanowires}

\author{C. M. Haapamaki}
\affiliation{Department of Engineering Physics, Centre for Emerging Device Technologies, McMaster University, Hamilton, Ontario, L8S 4L7, Canada}


\author{J. Baugh}
\affiliation{Institute for Quantum Computing, University of Waterloo, Waterloo, Ontario, N2L 3G1, Canada}
\affiliation{Department of Chemistry, University of Waterloo, Waterloo, Ontario, N2L 3G1, Canada}

\author{R. R. LaPierre}
\email{lapierr@mcmaster.ca}
\affiliation{Department of Engineering Physics, Centre for Emerging Device Technologies, McMaster University, Hamilton, Ontario, L8S 4L7, Canada}

\date{\today}

\begin{abstract}
InAs nanowires with Al$_x$In$_{1-x}$P or Al$_x$In$_{1-x}$As shells were grown on GaAs substrates by the Au-assisted vapour-liquid-solid (VLS) method in a gas source molecular beam epitaxy (GS-MBE) system. Core diameters and shell thicknesses were measured by transmission electron microscopy (TEM). These measurements were then related to selected area diffraction (SAD) patterns to verify either interface coherency or relaxation through misfit dislocations.  A theoretical strain model is presented to determine the critical shell thickness for given core diameters. Zincblende stiffness parameters are transformed to their wurtzite counterparts via a well known tensor transformation. An energy criterion is then given to determine the shell thickness at which coherency is lost and dislocations become favourable. 

\end{abstract}

\maketitle
\section{Introduction}\label{sec:intro}

III-V semiconducting nanowires are subject to great interest not only as building blocks for solar cells~\cite{Czaban2009}, detectors~\cite{Wei2009}, and quantum computing devices~\cite{Baugh2010}, but also as novel probes of exotic condensed matter phenomena.\cite{Mourik2012} In all of these applications, optimal device performance depends on the removal of surface states. The elimination of surface states reduces unwanted ionized impurity scattering, scattering from rough oxidized surfaces, electron-hole recombination, and carrier depletion due to surface traps. There are two methods generally used to passivate nanowire surfaces: chemical and structural. Chemical passivation of III-V nanowires is typically accomplished with ammonium polysulfide $(NH_4)_2S_x$~\cite{Suyatin2007,Tajik2012} or organic sulfide octadecylthiol (ODT).\cite{Sun2012} The effectiveness of chemical passivation declines with time so while chemical passivation is a good method to improve ohmic contact formation, it is not the best choice for long term passivation of the entire nanowire surface. The alternative to chemical passivation is structural passivation where a material with a higher band gap is grown around the nanowire core. Improved performance of InAs-InP core-shell nanowire field effect transistors (NWFET) has been shown over unpassivated InAs NWFETs.\cite{Dayeh2007} One disadvantage of structural passivation is that the shell must be etched selectively to form contacts to the core. Fortunately, selective etching is a common practice in III-V device processing so etching the shell is generally possible.\cite{Clawson2001}  

The main challenge in realizing core-shell structures is the strain (and dislocations) that result from the lattice mismatch. It has been shown that field effect mobility is reduced in core-shell structures that have undergone strain relaxation through the formation of dislocations.\cite{Kavanagh2011} To reduce the likelihood of strain relaxation, a suitable shell material must be chosen. In some cases the core and the shell can be lattice matched as in the case of GaAs-Al$_{0.52}$In$_{0.48}$P core-shell nanowires, eliminating the possibility of strain relaxation.~\cite{Chia2012} In the case of InAs cores however, there are no III-V materials available (excluding ternary antimonides) for lattice matching. As a result, care must be taken not to exceed the thickness beyond which the formation of dislocations becomes favourable. The traditional way of computing the critical thickness of two dimensional epilayers is by the Matthews model.\cite{Matthews1975} In recent years there have been a number of models that have addressed the issue of dislocations in the cylindrical core-shell geometry. These models typically have considered an isotropic material and calculated the energy required to form a dislocation.\cite{Gutkin2000,Colin2002,Kolesnikova2004,Ovidko2004,Ertekin2005,Aifantis2007,Fang2009} Some models have included the interactions between dislocations and strain fields.\cite{Ertekin2005} More specific critical thickness models have been developed for a variety of systems including Si-Ge~\cite{Trammell2008} and Al(In)GaN.\cite{Raychaudhuri2006} We have chosen to focus on the In containing ternary alloys AlInP and AlInAs for the shell material. InP shells on InAs cores proved to be difficult to achieve via GS-MBE so the Al-based ternary alloys were proposed as an alternative.\cite{Haapamaki2012} The addition of Al naturally promotes the formation of a shell by changing the growth mode from Au assisted VLS to step flow growth on the nanowire sidewall. Nanowires were grown in a GS-MBE system (SVT Associates). Group III species (In, Al) were supplied as monomers from effusion cells while group V species (As, P) were supplied as dimers cracked from hydrides. Details of the experimental parameters and nanowire characterization are described elsewhere.\cite{Haapamaki2012} 
\section{Model}\label{sec:model}

The strain model presented here was developed using the method of Trammell et al. where Si-Ge core-shell nanowires were studied.\cite{Trammell2008} This model was extended to describe the wurtzite structures observed in our InAs-AlInAs(P) nanowires as verified by TEM. In bulk form, all of the alloys considered here adopt the cubic zincblende crystal structure. As a result, we must convert the relevant parameters for the model from the tabulated zincblende values to their wurtzite counterparts. The in-plane equilibrium lattice constant is given by $a_{o} ^r = a_{ZB}/\sqrt{2}$ and the c-axis equilibrium lattice constant is given by $a_o ^z = \sqrt{8/3}\;a_o ^r$ where $a_{ZB}$ is the bulk zincblende lattice constant.\cite{De2010} The well known Martin transformation was used to obtain the stiffness tensor with results shown in Table~\ref{tab:zbwz}.~\cite{Martin1972} 
\begin{table*}
\caption{Non-zero components of the stiffness tensor $c_{ij}$. Shown are the tabulated zincblende~\cite{Vurgaftman2001} and calculated wurtzite values. Components of the stiffness tensor for ternary alloys Al$_x$In$_{1-x}$P and Al$_x$In$_{1-x}$As were computed using Vegard's law.\cite{Vurgaftman2001}}
\begin{center}
\begin{tabular*}{\textwidth}{@{\extracolsep{\fill}}lcccccccccc}\toprule
&\multicolumn{3}{c}{Zincblende}&\hspace{1cm}&\multicolumn{6}{c}{Wurtzite}\\
&\multicolumn{3}{c}{[GPa]}&\hspace{1cm}&\multicolumn{6}{c}{[GPa]}\\
\hline
 &$c_{11}$&$c_{12}$&$c_{44}$&\hspace{1cm}&$c_{11}$ & $c_{33}$ & $c_{12}$ & $c_{13}$ & $c_{44}$  & $c_{66}$ \\
InP\hspace{1cm} & 101.1 &	56.1  & 45.6  &\hspace{1cm}& 104.95 &131.90 &67.65	&40.70	&14.86	&18.65 \\
AlP\hspace{1cm}  & 188.3 &	67.1  &	36.9  &\hspace{1cm} &164.19	&156.70	&75.41	&82.90	&52.22	&44.39 \\
InAs\hspace{1cm} & 83.29 &	45.26 &	39.59 &\hspace{1cm}& 87.83	&110.72	&54.43	&31.54	&13.20	&16.70 \\
AlAs\hspace{1cm} & 119.9 &	57.5  &	56.6  &\hspace{1cm}&128.87	&153.77	&65.46	&40.57	&26.13	&31.71\\
\botrule
\end{tabular*}
\end{center}
\label{tab:zbwz}
\end{table*}

The Martin transformation is motivated by the strong similarity in tetrahedral coordination between the zincblende and wurtzite crystal structures. Locally the tetrahedra are rotated with respect to each other but otherwise identical. This allows the combination of a rotation and an internal strain compensation to estimate the wurtzite stiffness constants given the zincblende values.  The validity of the transformation has been verified by experiment in semiconductors that readily crystallize in both the zincblende and wurtzite structures such as ZnS.\cite{cline1967} In our study the nanowire cores were InAs, while the shell material was a ternary alloy of Al$_x$In$_{1-x}$As or Al$_x$In$_{1-x}$P. The lattice parameters and the stiffness constants for these materials were calculated using Vegard's law.\cite{Vurgaftman2001}

\begin{figure}
\includegraphics[scale=0.5]{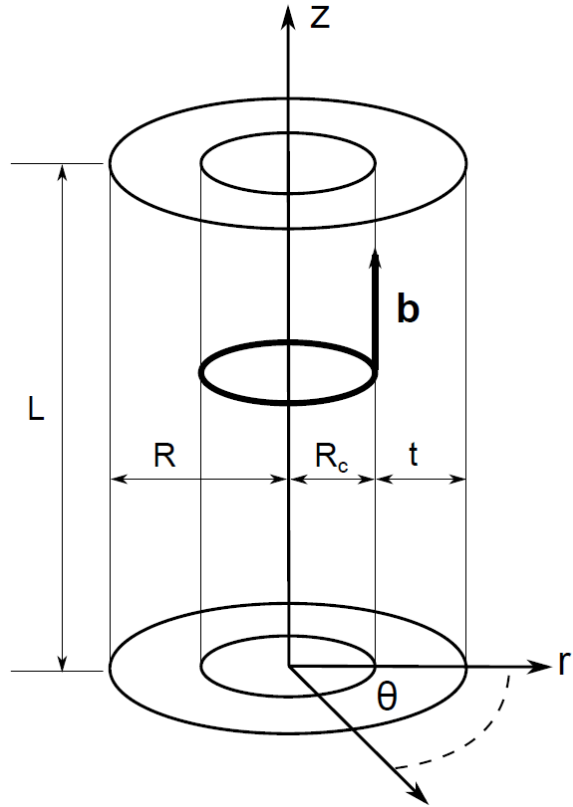}
\caption{Geometry of the core-shell nanowire of length $L$, radius $R$, core radius $R_c$, and shell thickness $t$ in the ($r,\theta,z$) coordinate system. Also shown is the location of a dislocation with Burgers vector $\bm b$.}
\label{fig:drawing}
\end{figure}

The nanowire geometry for this model consists of two coaxial cylinders (referred hereafter as core and shell) of different lattice constant and stiffness tensor. The strain was calculated through a two-step equilibration process. In Step 1, the core and shell were considered in their initial un-strained state and to be incoherent with each other. The reference state was then defined by establishing the condition of epitaxy where a single lattice constant is maintained throughout the system. For this model, the reference in-plane and z-axis lattice constants were chosen to be those of the core, $a ^{\mathrm{ref}}=a_r ^{c}$ and $a ^{\mathrm{ref}}=a_z ^c$. In Step 2 the system was allowed to elastically relax to the final state. This results in an expression for the total strain for $k=r,\theta,z$:

\begin{equation}
\label{eq:1}
\mathrm{e} _k ^i  = \frac{a_k ^c}{a_k ^i}\left ( \epsilon _k ^i - m_k ^i \right )
\end{equation}

\noindent where $m_k ^i=(a_k ^i - a_k ^c)/a_k ^c$ is the misfit strain from the reference state in Step 1, $\epsilon _k ^ i=(a_f ^i - a_k ^c)/a_k ^c$ is the displacement strain resulting from Step 2 where $a_f ^i$ is the final lattice constant, and $i$ refers to the core or shell. 

The displacement strains from Eq.~\eqref{eq:1} can be written in terms of the actual displacement $u_{r,\theta,z}$ in cylindrical coordinates:\cite{Timoshenko1951} 

\begin{align}
\epsilon _{r}  & = \frac{\partial u_r }{\partial r}\label{eq:e_f_urr}\\
\epsilon _{\theta}  &= \frac{u_r}{r}\label{eq:e_f_utt}\\
\epsilon _{z}  & = \frac{\partial u_z }{\partial z}\label{eq:e_f_uzz}
\end{align}

These displacements were solved by applying a series of boundary conditions appropriate to the core-shell nanowire geometry. The first boundary condition is a statement of mechanical equilibrium due to the absence of external loading:\cite{Timoshenko1951} 

\begin{align}
\frac{1}{r}\frac{\partial \sigma _r  r}{\partial r}-\frac{\sigma _{\theta}}{r}=0\label{eq:mecheq1}\\
\frac{\partial \sigma _z}{\partial z}=0\label{eq:mecheq3}
\end{align}

\noindent where the $\sigma _k$ are given by Hooke's law (See Reference~\cite{Trammell2008_2}):

\begin{align}
\sigma _r&=c_{11} \mathrm{e}_r +c_{12} \mathrm{e}_{\theta} +c_{12} \mathrm{e}_z \label{eq:s1}\\
\sigma _{\theta} &=c_{12} \mathrm{e}_r +c_{11} \mathrm{e}_{\theta} +c_{12} \mathrm{e}_z \label{eq:s2}\\
\sigma _z &=c_{12} \mathrm{e}_r +c_{12} \mathrm{e}_{\theta} +c_{11} \mathrm{e}_z \label{eq:s3}
\end{align}  

\noindent where $c_{ij}$ are the stiffness tensor elements from Table~\ref{tab:zbwz}. Substituting Eq.~\eqref{eq:s1}-\eqref{eq:s3} into Eq.~\eqref{eq:mecheq1}-\eqref{eq:mecheq3} we obtain the system of four displacement equations; $u_r$ and $u_z$ in the core and shell:

\begin{align}
\frac{\partial ^2 u_r }{\partial r^2}+\frac{1}{r}\frac{\partial u_r }{\partial r}-\frac{u_r }{r^2}=0\label{eq:disp1}\\
\frac{\partial ^2 u_z }{\partial z^2}=0\label{eq:disp2}
\end{align} 

\noindent Solutions to the displacement equations for the core and shell are:

\begin{align}
u_r & = \alpha  r+\frac{\beta }{r}\label{eq:dispsol1}\\
u_z &= \gamma  z + \phi \label{eq:dispsol2}
\end{align}

To obtain the eight coefficients $\alpha,\beta,\gamma,\phi$ for both core and shell in the above solution we apply boundary conditions arising from geometrical and physical constraints. In the nanowire core, as $r\rightarrow 0,\;\;u_r ^c \rightarrow \infty$ so $\beta ^c=0$. Continuity at the interfaces was maintained by requiring that the lattice constants were continuous across the interface. In the z-direction, this eliminates $\phi$ and requires $\gamma ^c = \gamma ^s$. The interface continuity is further specified by maintaining continuity in displacement and stress across the interface:

\begin{align}
&u_r ^c(r=R_c) - u_r ^s(r=R_c) = 0 \label{eq:contdisp}\\
&\sigma _r ^c (r=R_c) - \sigma _r ^s (r=R_c) = 0 \label{eq:contsigma}
\end{align}

Since the nanowire is in mechanical equilibrium, the net force on the surfaces must be zero which is expressed as:

\begin{align}
&\int _0 ^{R_c} \int _0 ^{2\pi}\sigma _z ^c r\,drd\theta - \int _{R_c} ^{R} \int _0 ^{2\pi}\sigma _z ^s r\,drd\theta=0\label{eq:nosurfFz}\\
&\sigma _r ^s (r=R_c) = 0\label{eq:nosurfstress}
\end{align} 

The resulting expressions for displacement are obtained through a lengthy solution to the system of four algebraic equations, Eq.~\eqref{eq:contdisp}-\eqref{eq:nosurfstress}, to determine the four remaining coefficients $\alpha ^c,\gamma ^c,\alpha ^s,\beta ^s$ (not shown here due to length). 

The limits of coherency were obtained by comparing the strain energy and the energy required to form a dislocation. To determine the strain energy, the displacement solutions were substituted back into the expressions for stress in Eq.\eqref{eq:s1}-\eqref{eq:s3} and in the expressions for strain in Eq.\eqref{eq:e_f_urr}-\eqref{eq:e_f_uzz}. These expressions were combined to compute the strain energy as follows:

\begin{align}
U^c = \frac{1}{2}\int _0 ^{L}dz\int _0 ^{R_c}dr\int _0 ^{2\pi} d\theta \;\sigma _k ^c \mathrm{e}_k ^c r \label{eq:coreenergy}\\
U^s = \frac{1}{2}\int _0 ^{L}dz\int _{R_c} ^{R}dr\int _0 ^{2\pi} d\theta \;\sigma _k ^s \mathrm{e}_k ^s r \label{eq:shellenergy}
\end{align}

\noindent where summation over $k=r,\theta,z$ is implied. In this model we only consider an edge dislocation where an extra plane of atoms is inserted in the shell in the $[0001]$ direction. This is the only dislocation we observed experimentally.\cite{Haapamaki2012} The formation energy for such a dislocation is:~\cite{Chou1962} 

\begin{equation}\label{eq:disenergy}
W=K \bm{b}^2 a \left ( \ln \frac{8a}{r_o}-2\right )
\end{equation}

\noindent where $\bm{b}$ is the burgers vector of the dislocation, which in this case is equal to the z-axis lattice constant of the shell. $a$ is the radius of the dislocation set equal to the core radius $R_c$, and $r_o$ is the cutoff radius that eliminates a mathematical singularity at the core of the dislocation. Eq.~\eqref{eq:disenergy} takes into account the core energy of the dislocation. The cutoff radius is set to $\bm{b}/4$ for semiconductors.\cite{vandermerwe1991} $K$ is the energy factor and is determined from the elastic constants:

\begin{equation}\label{eq:efactor}
K = \left ( \bar{c}_{13}+c_{13}\right ) \left [ \frac{c_{44}\left (\bar{c}_{13}-c_{13}\right )}{c_{33}\left (\bar{c}_{13}+c_{13}+2c_{44}\right )} \right ] ^{1/2}
\end{equation}

\noindent where $\bar{c}_{13}=\sqrt{(c_{11}c_{33})}$. To determine the point at which a dislocation is formed, we determine the geometry (core radius $R_c$ and shell thickness $t$) at which the inequality $U^c + U^s > W$ occurs. 

\section{Results}\label{sec:results}

\begin{figure}
\includegraphics[width=\textwidth]{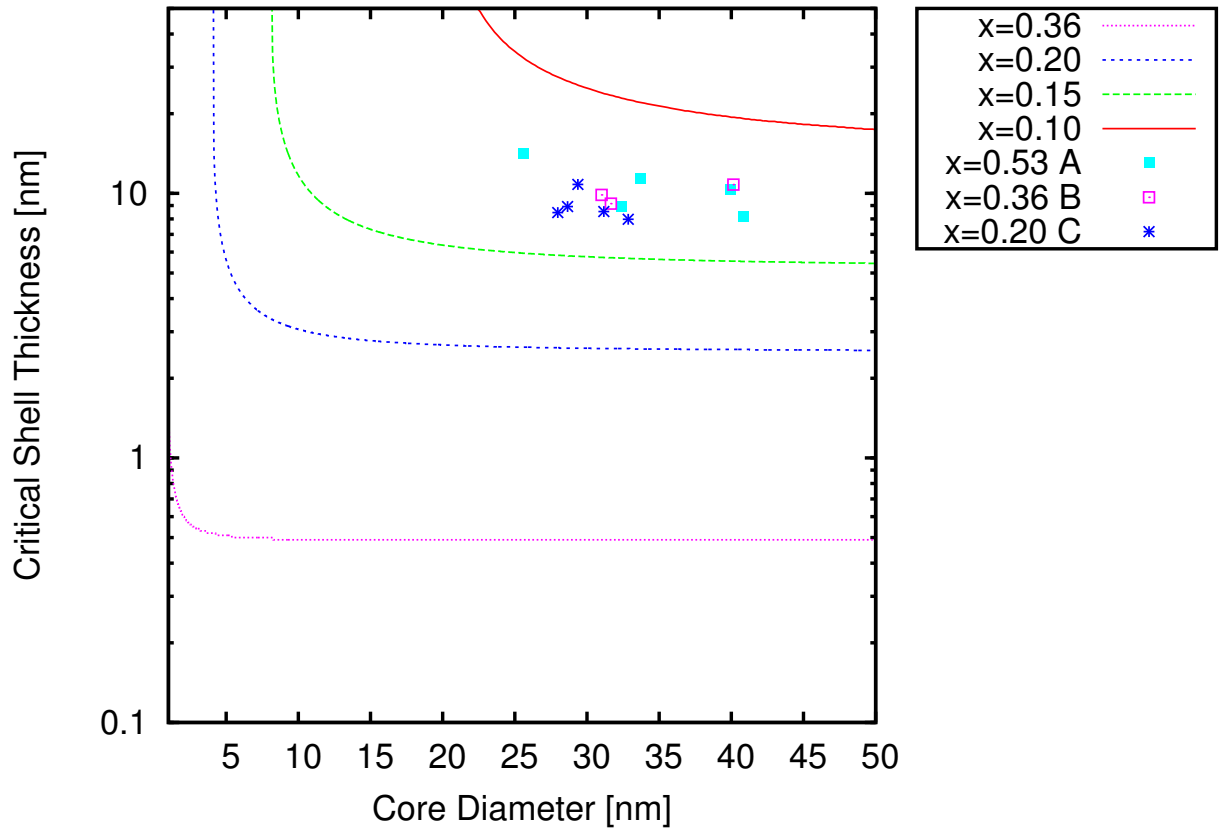}
\caption{Plot of the critical Al$_x$In$_{1-x}$As shell thickness as a function of InAs core diameter for different alloy parameters, $x$. Also shown are data points from nanowires grown previously denoted A, B, and C.}\label{fig:critAlInAs}
\end{figure}

\begin{figure}
\includegraphics[width=\textwidth]{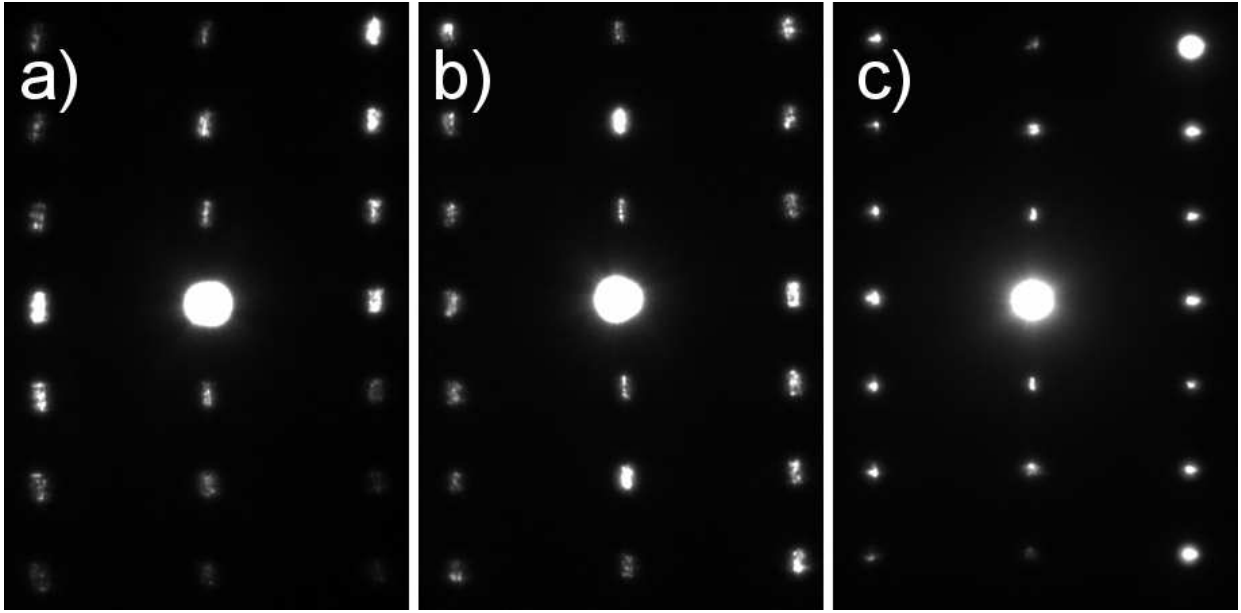}
\caption{SAD pattern of InAs-Al$_{x}$In$_{1-x}$As core-shell nanowires along the $[2\bar{1}\bar{1}0]$ zone axis for a) $x=$ 0.53, 3.6\% mismatch, b) $x=$ 0.36, 2.5\% mismatch, and c) $x=$ 0.20, 1.3\% mismatch. A, B, C from Figure~\ref{fig:critAlInAs} correspond to a), b) and c) in this figure.}\label{fig:sadAlInAs}
\end{figure}

\begin{figure}
\includegraphics[width=\textwidth]{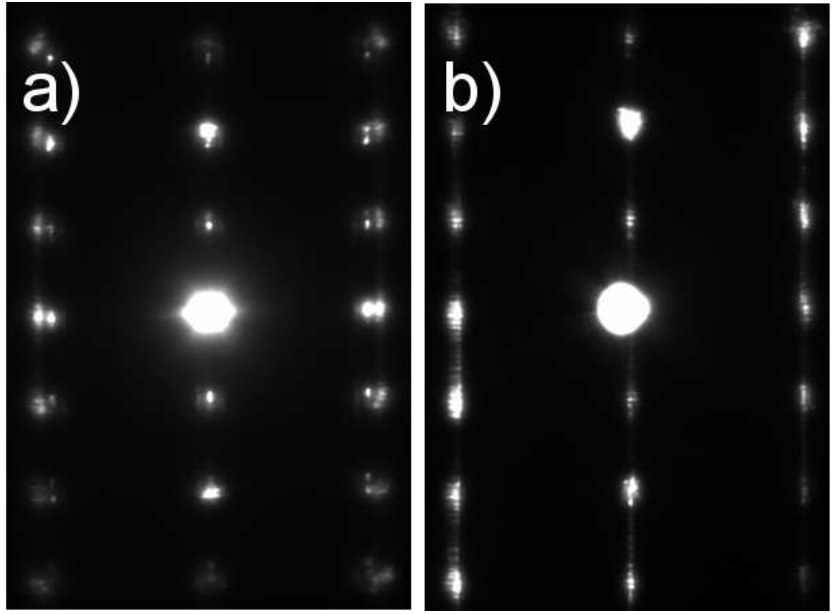}
\caption{SAD pattern of InAs-Al$_{x}$In$_{1-x}$P core-shell nanowires along the $[2\bar{1}\bar{1}0]$ zone axis for a) $x=$ 0.53, 7.1\% mismatch, b) $x=$ 0.36, 5.8\% mismatch}\label{fig:sadAlInP}
\end{figure}

Figure~\ref{fig:critAlInAs} shows the calculated critical thickness for InAs-Al$_x$In$_{1-x}$As core-shell nanowires. As the core radius increases to infinity, the critical thickness becomes constant, meaning the model reduces to the thin film case as expected. The results in Figure~\ref{fig:critAlInAs} were compared with experimental data. The experimental data points in Figure~\ref{fig:sadAlInAs} indicate the shell thickness and core diameter measured by HRTEM for a number of InAs-Al$_x$In$_{1-x}$As core-shell nanowires with nominal composition $x=$ 0.53 (sample A), 0.36 (sample B) and 0.20 (sample  C) as reported previously.\cite{Haapamaki2012} The measured shell thickness and core diameter was in the range of 8-14 nm and 25-40 nm, respectively. In Figure~\ref{fig:sadAlInAs}, SAD images are shown from InAs-Al$_x$In$_{1-x}$As core-shell nanowires.\cite{Haapamaki2012} Moving from a)-c) we show a decrease in alloy fraction and a subsequent decrease in spot splitting. Spot splitting and spot broadening are associated with strain relaxation, which in this case is due to the epitaxial mismatch between the InAs core and the Al$_x$In$_{1-x}$As shell. Similarly, in Figure~\ref{fig:sadAlInP} we show SAD patterns of nanowires with Al$_x$In$_{1-x}$P shells on InAs cores where larger relaxation is evident by further spot splitting. For all combinations of shell thickness and core diameter, nanowires in Figure~\ref{fig:sadAlInP} well exceeded the calculated critical thickness and will not be discussed further. Instead, we focus on the InAs-Al$_x$In$_{1-x}$As core-shell nanowires.

The shell thickness (8-14 nm) for samples A and B well exceeded the corresponding theoretical critical thickness curves for $x=$ 0.53 and $x=$ 0.36 nm. The critical thickness in these cases is below 1 nm in the diameter range of 25-40 nm. Hence, the shells in these samples are expected to relax as observed in Figure~\ref{fig:sadAlInAs}. Dislocations were also observed directly in HRTEM as described previously.\cite{Haapamaki2012} 

Experimentally, no dislocations were observed in nanowires with $x=$ 0.20 (sample C) consistent with the sharp spots in Figure~\ref{fig:sadAlInAs}c). We concede however that finding a single dislocation in a nanowire that is 1 $\mu$m in length is not an easy task by HRTEM. Energy dispersive x-ray analysis (EDX) on these nanowires found that the actual composition of the shell in sample C nanowires varied from $x=$ 0.10 to 0.15, somewhat below the nominal value of $x=$ 0.20. Due to the proximity of the experimental data for sample C to the $x=$ 0.10 theoretical curve in Figure~\ref{fig:critAlInAs}, we might expect the shells of these samples to be coherent as observed in Figure~\ref{fig:sadAlInAs}. Overall, these data suggest that Al$_x$In$_{1-x}$As shells with $x<$ 0.15 and thickness below 15 nm on InAs cores (25-40 nm diameter) are coherent and free of misfit dislocations. 

\section{Summary}

We have outlined a model to calculate the critical thickness for core-shell InAs-AlInAs(P) nanowires. Comparing the strain energy and the dislocation formation energy, we were able to determine the point at which the system underwent strain relaxation via the insertion of a misfit dislocation. Extending the model by Trammell~\cite{Trammell2008}, and using the transform by Martin~\cite{Martin1972} we have taken into account the wurtzite crystal structure of our nanowires. Comparisons with SAD patterns of InAs-AlInAs(P) core-shell nanowires with varying Al alloy parameter showed different degrees of strain accommodation due to lattice mismatch. The model predicted strain relaxation in all samples that showed SAD spot splitting. The Al$_{0.20}$In$_{0.80}$As shell showed no SAD spot splitting, consisten with the model predictions when the actual Al alloy fraction measured by EDX is considered.

\begin{acknowledgments}
Financial assistance from the Natural Sciences and Engineering Research Council of Canada is gratefully acknowledged.  Assistance with the TEM by Fred Pearson and with the MBE growths by Shahram Tavakoli are gratefully acknowledged. We also acknowledge discussions with Elizabeth Dickey regarding our strain analysis.
\end{acknowledgments}

\bibliographystyle{aipnum4-1}
\bibliography{haapamaki}

\end{document}